\documentclass[reprint, superscriptaddress, nofootinbib, aps, preprintnumbers ,twocolumn]{revtex4-2}
\usepackage[utf8]{inputenc}
\usepackage{amsfonts}
\usepackage{amsmath,amssymb}
\usepackage{bbm}
\usepackage{soul}
\usepackage{bm}
\usepackage{color}
\usepackage{slashed}
\usepackage[most]{tcolorbox}
\usepackage[toc,page]{appendix}
\usepackage{graphicx}
\usepackage{placeins}
\usepackage{caption}
\usepackage{xcolor}
\usepackage{subfigure}
\usepackage{booktabs}
\usepackage{microtype} 
\usepackage{comment}
\usepackage{physics}
\usepackage{float}
\usepackage{bbold}
\usepackage{ragged2e}
\usepackage{dcolumn}
\DeclareGraphicsExtensions{.pdf,.png,.jpg}

\usepackage[colorlinks]{hyperref}
\AtBeginDocument{%
  \hypersetup{
    citecolor=blue,
    linkcolor=blue,   
    urlcolor=blue}}

\definecolor{airforceblue}{rgb}{0.36, 0.54, 0.66}
\definecolor{steelblue}{rgb}{0.27, 0.51, 0.71}
\definecolor{amber}{rgb}{1.0, 0.49, 0.0}
\definecolor{darkgreen}{rgb}{0.0, 0.5, 0.0}
\definecolor{amber}{rgb}{1.0, 0.49, 0.0}

\DeclareMathAlphabet{\mathpzc}{OT1}{pzc}{m}{it}

\graphicspath{{Figures/}}
\allowdisplaybreaks[1]
\def\simg{{\ \lower-1.2pt\vbox{\hbox{\rlap{$>$}\lower6pt\vbox{\hbox{$\sim$}}}}\ }}
\def\siml{{\ \lower-1.2pt\vbox{\hbox{\rlap{$<$}\lower6pt\vbox{\hbox{$\sim$}}}}\ }}

\makeatletter \@addtoreset{equation}{section} \makeatother

\newcommand*{\diff}[1]{\text{d}#1}

\newcommand{\IN}[1]{\int_\Omega\diff{#1}\, }

\begin{document}
\preprint{IFT-UAM/CSIC-23-139}

\title{Gravity-Matter Sum Rules \\ in models with a single extra-dimension}

\author{A. de Giorgi}\email{arturo.degiorgi@uam.es}
\affiliation{%
 Departamento de F\'isica Te\'orica and Instituto de F\'isica Te\'orica UAM/CSIC,\\
Universidad Aut\'onoma de Madrid, Cantoblanco, 28049, Madrid, Spain
}
\affiliation{%
Department of Physics \& Laboratory for Particle Physics and Cosmology,\\
Harvard University, Cambridge, MA 02138, USA
}%

\author{S.~Vogl}\email{stefan.vogl@physik.uni-freiburg.de}
\affiliation{%
Albert-Ludwigs-Universität Freiburg, Physikalisches Institut\\
Hermann-Herder-Str. 3, 79104 Freiburg,
Germany
}%
\begin{abstract}
  We prove a set of sum rules needed for KK-graviton pair production from matter in orbifolded extra-dimensional models. The sum rules can be found in full generality by considering the properties of solutions to the Sturm-Liouville problem, which describes the wave functions and the masses of the KK-gravitons in four dimensions.
  They ensure cancellations in the amplitudes of the processes mentioned above which considerably reduce their growth with $s$ in the high-energy limit. This protects extra-dimensional theories from the low-scale unitarity problems that plague other theories with massive spin-2 particles. 
  We argue that such relations are valid for a broader category of models thus generalizing our previous results that were limited to the large $\mu$ limit of the Randall-Sundrum model.
\end{abstract}
\maketitle
\tableofcontents
\begin{section}{Introduction}

As is well known the longitudinal mode of the polarization tensor of massive spin-2 particles is proportional to $E^2/m^2$ where $E$ is the energy of the particle and $m$ its mass. This leads to a strong growth of amplitudes with these particles as external states with the Mandelstam $s$ in the Fierz-Pauli theory~\cite{Fierz:1939ix}. Consequently,  perturbative unitarity breaks down at a significantly lower energy scale than the effective theory suggests. For spin-2 scattering the amplitude grows as $s^5$ while the production of pairs of spin-2 particles from matter fields grows as $s^3$. 
Due to the clear connection between gravity and spin-2 fields, there is a great interest in curing this behaviour since this might allow the construction of a theory of massive gravity which could help to address various cosmological questions. At present, the best-behaved theory with a single massive spin-2 particle features polynomial interaction terms for the spin-2 particle that reduce the growth in scattering to $s^3$ \cite{Arkani-Hamed:2002bjr,deRham:2010kj,Hinterbichler:2011tt}. 

The situation is different in extra-dimensional theories where massive spin-2 particles arise as degrees of freedom in an effective theory from the Kaluza-Klein~(KK) decomposition of higher-dimensional gravity. In this case, the fundamental scale of the theory is given by the higher-dimensional Planck mass. Therefore, a lower breakdown scale, as implied by a growth of the amplitudes faster than $s$, is not expected. It is avoided by a subtle cancellation between the contributions of the full KK tower which restores perturbative unitarity of tree level amplitudes up to the Planck scale  \cite{Schwartz:2003vj,SekharChivukula:2019yul,SekharChivukula:2019qih,Chivukula:2020hvi,Bonifacio:2019ioc,deGiorgi:2020qlg,Chivukula:2022tla}. To achieve this cancellation the KK-gravitons need to fulfil several intricate consistency conditions that can be expressed as sum rules for masses and the coefficient of the KK-graviton vertices.

In this work, we discuss how the sum rules for graviton pair production can be proven in a general and mathematically exact way by studying the properties of the Sturm-Liouville~(SL) equation. The explicit connection between SL equation and scattering amplitudes involving KK gravitons was first carried out in Ref.~\cite{SekharChivukula:2019yul}. In such reference and in Refs.~\cite{SekharChivukula:2019qih, Chivukula:2020hvi} the properties of the SL equation are used to show how amplitudes in KK-scattering in large-$s$ limit vanish. On the same lines, we apply such a technique to the case of graviton-matter annihilation. This generalizes the results of our earlier work \cite{deGiorgi:2020qlg} were a proof of the sum rules in the large $\mu$ limit of the RS-model was given \footnote{Another work that studies this generalization and also considers bulk matter fields appeared simultaneously to this work in  Ref.~\cite{Chivukula:2023sua}.}. 

For illustration, we demonstrate how they cancel the leading powers in $s$ of an expansion of the amplitudes in~$s$
\begin{equation}
    \mathcal{M}(s)=\mathcal{M}^{(3)}s^3+\dots\,,
\end{equation}
for KK-graviton pair production
explicitly in RS.
These cancellations are of great interest for phenomenological studies. For example, spin-2 particles have been considered as mediators to the dark sector or directly as dark matter particles \cite{Lee:2013bua,Rueter:2017nbk,Babichev:2016bxi,Kraml:2017atm,Folgado:2019sgz,deGiorgi:2021xvm,deGiorgi:2022yha,Gill:2023kyz,Gonzalo:2022jac,Anchordoqui:2022svl}. For these studies it is of crucial importance how the amplitudes that control the cross sections behave at high momentum transfer.

 The paper is organized as follows. First, we discuss some general properties of compactifications and the Sturm-Liouville equations that are then used to derive a set of sum rules for the masses and integrals of the wave functions. In Sec.~\ref{sec:physics_application} we apply these to scattering processes in the RS model and demonstrate that they are sufficient to ensure the cancellations of the most offending contributions to the amplitude and bring their growth with energy down. However, they are not sufficient to cancel the spurious growth completely and need to be supplemented by compactification-dependent sum rules for the radion contribution to achieve the full cancellation.
Finally, we summarize our results and provide a brief outlook in Sec.~\ref{sec:summary_and_outlook}.

\end{section}

\begin{section}{Compactification and Sturm-Liouville equation}

\subsection{Compactification}
Our primary focus is on orbifolded 5D theories of gravity with an extra compact dimension of size $R$. More specifically, the 5D space-time is compactified under an $S^1/\mathbb{Z}^2$ orbifold symmetry yielding a 1D bulk bounded by two 4-dimensional (4D) branes located at $y=0$ and $y=\pi R$, where $y$ is the coordinate of the 5th-dimension.
We consider Einstein-Hilbert gravity in five dimensions
    \begin{align}
    \label{eq:RS_Action}
            & S_{\text{bulk}} = \frac{M_5^3}{2}\int \diff{}^4x \int\limits_{-\pi R}^{\pi R} \diff{y} \sqrt{g}(\mathcal{R}-2\Lambda_B )\,,
    \end{align}
where $g$ is the determinant of the 5D metric, $\mathcal{R}$ the Ricci scalar, $M_5$ the 5D Planck mass, and  $\Lambda_B$ denotes the vacuum energy of the bulk.
Regarding the matter content, we consider models where matter fields are localized on the branes, i.e. on the boundaries of the 5th dimension.

The gravitational particle content is obtained by expanding the metric around the 5D-Minkowski metric
\begin{equation}
\label{eq:metric-expansion}
    g_{MN}=\eta_{MN}+\kappa h_{MN}\,,
\end{equation}
where $h_{MN}$ is the metric perturbation  $\kappa\equiv 2/M_5^{3/2}$.
In the suitable coordinates, the relevant components of the metric perturbation $h_{MN}$ are its tensorial, $h_{\mu\nu}$, and scalar, $r$, ones; they are typically dubbed \textit{graviton} and \textit{radion}, respectively. Five-dimensional fields can be expanded via the so-called \textit{Kaluza-Klein(KK)-decomposition} in terms of an orthonormal basis, $\{\psi_n(y)\}_n$,  of the compactified dimension. For $h_{\mu\nu}(x,y)$ it amounts to
\begin{equation}
    h_{\mu\nu}(x,y)=\dfrac{1}{\sqrt{R}}\sum\limits_{n=0}^\infty h^{(n)}_{\mu\nu}(x)\psi_n(y)\,,
\end{equation}
In the effective 4D theory, this procedure generates an infinite tower of 4D gravitons.
The expansion of the metric of Eq.~\eqref{eq:metric-expansion} can be conveniently chosen such that the 4D gravitons are already in the mass basis and have canonical kinetic and mass terms.
As the specific value of $R$ is not relevant to our discussion, we work with $y$ in units of $R$, which effectively amounts to setting $R=1$, and thus $y\in[-\pi,\pi]$.

The ansatz for the metric can be parametrized as~\cite{Giudice:2016yja}
\begin{equation}
\label{eq:metric}
    \diff s^2= e^{-2k|y|}\left(\eta_{\mu\nu}\diff{x}^\mu
\diff{x}^\nu - e^{6lk|y|} \diff y^2\right)\,,
\end{equation}
where $\eta_{\mu\nu}=\text{diag}(1,-1,-1,-1)$ is the Minkowski metric and $k\sim\sqrt{-\Lambda_B}$.
This ansatz interpolates between several popular models, i.e. LED~($k=0$)~\cite{Arkani-Hamed:1998jmv}, RS~($l=1/3$)~\cite{Randall:1999ee} and CW~($l=0$)~\cite{Giudice:2016yja}.
The corresponding equation of motion for the graviton wave function along the direction of the fifth dimension, $\psi_n$, reads
\begin{equation}
\label{eq:generic-metric-eq}
    \left[\partial_y^2-\frac{9}{4}(1+l)^2k^2+e^{6lk|y|}m_n^2\right]e^{-\frac{3}{2}(1+l)k|y|}\psi_n(|y|)=0 \\,,
\end{equation}
where $m_n$ is the mass of the graviton $h_{\mu\nu}^{(n)}$.
This is a particular case of a Sturm-Liouville (SL) equation as expected given that the Einstein-Hilbert action only contains up to two derivatives and all homogeneous second-order ordinary differential equations can be brought to SL form. 

To ease readability of the main text, we briefly review some of them in the next section.

\subsection{The Sturm-Liouville Equationn}
\subsubsection{General Properties}
The SL equation is a real second-order linear ordinary differential equation with interesting properties and many applications in physics. In the following, we briefly recapitulate some of the basics that matter for our discussion. For a more systematic presentation of its mathematical properties and the proofs see for example Ref.~\cite{SL-book}. In its most general form, it reads
\begin{equation}
\label{eq:SL}
    -\frac{1}{\diff y}\left[p(y)\frac{\diff \psi}{\diff y}\right] + q(y)\psi = \lambda r(y)\psi \,.
\end{equation}
The function $r(y)$ is sometimes referred to as the \textit{weight} and $\lambda$ as the \textit{eigenvalue} of the equation.
The SL equation together with boundary conditions for the solution constitutes the so-called \textit{SL problem}. The eigenvalue is generically not specified, and it is part of the SL problem to find suitable eigenvalues for which non-trivial solutions exist.
If $p(y)$ and $r(y)>0$ while $p(y),\,p'(y),\,q(y),\,r(y)$ are continuous in the interval $[a,b]$, and the boundary conditions are given in the form
\begin{equation}
\label{eq:regular-cond}
    \alpha_1 \psi(a)+\alpha_2 \psi'(a)=0\,,\qquad  \beta_1 \psi(b)+\beta_2 \psi'(b)=0\,,
\end{equation}
the SL problem is said to be \textit{regular}. Regular SL problems satisfy the following properties:
\begin{enumerate}
    \item There exist an infinite countable number of real eigenvalues $\{\lambda_n\}_{n=0}^\infty$ that can be ordered such that
    \begin{equation}
        \lambda_0<\lambda_1<\dots< \lambda_n <\dots <\infty\,.
    \end{equation}
    \item For each eigenvalue $\lambda_i$ there exists a unique eigenfunction $\psi_i(y)$, up to rescalings.
    \item The solutions form an orthonormal basis of the Hilbert space $L^2([a,b],r(y)\diff{y})$, that is
    \begin{equation}
    \label{eq:othonormal-psi}
        \braket{\psi_i}{\psi_j}\equiv \int_a^b\diff{y}\,r(y) \psi_i(y)\psi_j(y) = \delta_{ij}\,.
    \end{equation}
\end{enumerate}
This also implies that any function, $f(y)$ can be expanded within the interval $(a,b)$ as
\begin{equation}
\label{eq:completeness}
    \begin{split}
        &f(y)\equiv \ket{f}=\sum\limits_{n=0}^\infty \ket{\psi_n}\braket{\psi_n}{f}=\sum\limits_{n=0}^\infty c_n \psi_n(y)\,,\\
        &\text{where} \quad c_n = \int_a^b \diff{y}\, r(y) \psi_n(y) f(y) \,.
    \end{split}
\end{equation}
Such information can be conveniently encoded in functional representation via the completeness relation 
\begin{equation}
\label{eq:delta-relation}
    \delta(y-y')=\sum\limits_{k=0}^\infty r(y)\psi_k(y)\psi_k(y') \,.
\end{equation}
Finally, one can show that, for $q=0$, if $\psi_i$ is a solution of the SL equation, then $p(y)\partial_y \psi_i(y)$ is also a solution with $r\to 1/p$ and $p\to 1/r$. This implies orthogonality relations among the derivatives of $\psi_i$ of the form
\begin{equation}
\label{eq:othonormal-derpsi}
    \int_a^b \diff{y}\, p(y) \left(\partial_y\psi_i\right)\left(\partial_y\psi_j\right) = \lambda_i \delta_{ij} \,.
\end{equation}

Given the above-mentioned properties, the SL equation can be recast as an eigenvalue problem (from which the name of $\lambda$ derives) of a linear operator $L$, such that
\begin{equation}
\label{eq:L}
    L[\psi_i](y)=\lambda_i \psi_i(y)\,,
\end{equation}
with
\begin{equation}
    L[f](y)\equiv -\dfrac{1}{r(y)}\left(\dfrac{\diff}{\diff{y}}\left[p(y)\dfrac{\diff f}{\diff{y}} \right]+q(y)f\right)\,.
\end{equation}
The SL operator is self-adjoint within the interval $[a,b]$. In the following, we will write $\Omega$ and $\partial\Omega$ to indicate the domain of integration and its boundary, respectively.

\subsubsection{Useful Definitions and basic properties}
The effective couplings involving the KK-tower of gravitons can be derived by integrating out the 5th dimension. They correspond to n-points integrals containing either no or two derivatives of $\psi_i$. For this reason, it is useful to study the $n-$point integrals
\begin{align}
\label{eq:def-a}
    a_{i_1,i_2,\dots,i_n} &\equiv \int_\Omega \diff y \ r(y) \psi_{i_1}\psi_{i_2}\psi_{i_3} \dots \psi_{i_n} \,,\\
    \label{eq:def-b}
    b_{i_1,i_2,\dots,i_n} &\equiv \int_\Omega \diff y \ p(y)\left(\partial_y\psi_{i_1}\right)\left(\partial_y\psi_{i_2}\right)\psi_{i_3} \dots \psi_{i_n} \,,
\end{align}
where the indices $i_1,i_2,\dots,i_n$ will indicate the number of the $n$ gravitons involved in the interaction. 
Notice that the first two indices of $b_{i_1,i_2,\dots,i_n}$ commute between themselves, but not with the others and vice versa.
With Eq.~\eqref{eq:othonormal-psi} and \eqref{eq:othonormal-derpsi} one immediately finds that for $n= 2$ the integrals are given by
\begin{align}
    &a_{ij}=\delta_{ij}\,,
    &b_{ij}=\lambda_i\delta_{ij}\,.
\end{align}
The $a_{i_1,\dots,i_n}$ and $b_{i_1,\dots,i_n}$ coefficients are not fully independent one from another. They can be related by integration by parts via Eq.~\eqref{eq:SL}. In fact, for any function $f(y)$ we have that
\begin{equation}
    \begin{split}
        \int_\Omega \diff{y} \ p(y) \partial_y f \partial_y \psi_i =&\left.p(y) f \partial_y \psi_i\right|_{\partial\Omega}+ \lambda_i \int \diff{y} \ r(y) f\psi_i\\
        &+ \int \diff{y} \ q(y) f\psi_i \,.
    \end{split}
\end{equation}

The above relation can be further simplified in the case $q=0$ and $\partial_y\psi_i|_{\partial\Omega}=0$. By choosing $\psi_i = \psi_{i_1}$ and $f(y)=\psi_{i_2}\dots \psi_{i_n}$, one obtains
\begin{equation}
\label{eq:a->b}
    \lambda_{i_1}a_{i_1,i_2,\dots,i_n} = \sum\limits_{j\in{i_2,\dots,i_n}}b_{i_1,j,X_j}\,,
\end{equation}
where $X_j$ includes all the indices except $i_1$ and $j$.
For $n=2$ this recovers the orthogonality relations. The first non-trivial case is for $n=3$. It reads
\begin{equation}
\label{eq:general-sturm}
    \lambda_i a_{ijk}=b_{ijk}+b_{ikj} \ ,
\end{equation}
which implies
\begin{equation}
\label{eq:b-from-a}
    b_{ijk}=\frac{1}{2}a_{ijk}\left(\lambda_i +\lambda_j-\lambda_k\right)\ .
\end{equation}
The above equation shows that for $n=3$ all $b$-coefficients can be unambiguously traded for $a$-coefficients. This special case allows the translation of many of the results for the sum rules of one type of coefficient to the other and vice versa.

In the next section, we will make use of such relations and definitions to prove sum rules involving $a$- and $b$-coefficients that will be relevant for the physics involving the scattering of gravitons and gravitons pair production.
\begin{subsection}{SL Sum Rules}
\label{sec:linear}
For convenience, the most relevant sum rules and relations needed for the amplitudes are collected in Table~\ref{eq:sumrules}.
\begin{table*}[tbh]
\centering
\begin{tabular}{lr}
\toprule
Sum Rule & $\mathcal{M}$ \\
\midrule
$\sum\limits_{k=0}^\infty\psi_k(\pi) a_{ijk}=\psi_i(\pi)\psi_j(\pi)$ & $\mathcal{M}^{(3)}$\\
$\sum\limits_{k=0}^\infty\psi_k(\pi)\lambda_k a_{ijk}=\psi_i(\pi)\psi_j(\pi)(\lambda_i+\lambda_j)$& $\mathcal{M}^{(2)}$\\
$\sum\limits_{k=0}^\infty\psi_k(\pi)b_{kij}=\lambda_i \psi_i(\pi)\psi_j(\pi)$ & $\mathcal{M}^{(2)}$\\
$\sum\limits_{k=0}^\infty \psi_k(\pi)b_{ijk}=0$ &$\mathcal{M}^{(2)}$\\
\midrule
$\sum\limits_{k=1}^\infty \dfrac{\psi_k(\pi)}{\lambda_k}a_{ijk}(\lambda_i-\lambda_j)^2=(\lambda_i+\lambda_j)\left[\psi_i(\pi)\psi_j(\pi)+\psi_0^2\delta_{ij}\right]-6 \psi_r b_{ijr}$ & $\mathcal{M}^{(2)}$\\
$ \sum\limits_{k=1}\psi_k(\pi) \dfrac{b_{kir}}{\lambda_k}=A_\pi^{-2}\psi_r\psi_i(\pi)-\psi_r a_{irr}$ & $\mathcal{M}^{(2)}$\\
\midrule
$b_{ijk}=\frac{1}{2}a_{ijk}\left(\lambda_i +\lambda_j-\lambda_k\right)$& -\\
\bottomrule
\end{tabular}
\caption{
Sum Rules and relations needed for the cancellations and highest order in $\mathcal{M}$ at which they appear. The first block includes sum rules involving solely the gravitons and their generality extends beyond the RS model. In the second block, two extra sum rules involving the radion in the RS model are given. The last relation at the bottom is useful to convert from $a$- to $b$- coefficients. It is also general.} 
\label{eq:sumrules}
\end{table*}
 The diagrams that involve such sums are shown in Fig.~\ref{fig:linear-diagrams}.
 \begin{figure*}[t]
    \centering
    \includegraphics[width=0.75\textwidth]{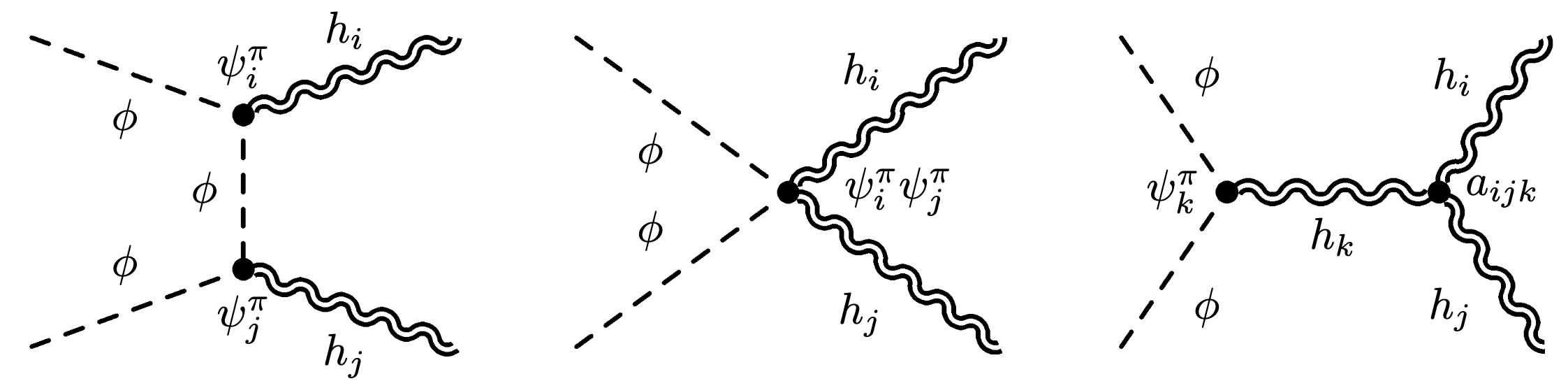}
    \caption{Representative set of diagrams contributing to the highest $\mathcal{M}^{(n)}$ for gravitons pair-production from matter. For convenience, we have used the notation $\psi_i^\pi\equiv \psi_i(\pi)$.}
    \label{fig:linear-diagrams}
\end{figure*}
As can be seen, the graviton wave functions enter in the total amplitude either localized on the brane or at most linearly via $a_{ijk}$ when integrated over the bulk. Since we have already in mind the physical application, we will consider $\partial\Omega=\{0,\pi\}$, but the results only require the constant parts to be evaluated on the boundary, i.e. $\psi_i(\partial\Omega)$, $(\partial_y\psi_i)(\partial\Omega)$. 

The main task is to compute sum rules involving $\sum_k \psi_k(\pi)a_{ijk}$~\footnote{Notice that all the sum rules we derive for $y=\pi$ equally apply to $y=0$.} and related.
  Applying the Liouville operator, one finds the general relation $(\alpha\geq0)$ 
\begin{equation}
\label{eq:linear-master-eq}
    \sum\limits_{n=0}^\infty \psi_n(\pi)\lambda_n^\alpha\left(\IN{y}r(y)\psi_n(y)f(y)\right) = \left.L^\alpha[f(y)]\right|_{y=\pi}\,.
\end{equation}
Taking $f(y)=\psi_i\psi_j$ and $\alpha=0$, leads to
\begin{align}
\label{eq:lin-0-sr}
    &\sum\limits_{k=0}^\infty\psi_k(\pi) a_{ijk}=\psi_i(\pi)\psi_j(\pi)\,.
\end{align}
This result is completely general, independently of $q(y)$ or on the boundary conditions. As for the quadratic sum rules, by dimensional analysis, Eq.~\eqref{eq:lin-0-sr} is relevant to cancel the leading order, $\mathcal{M}^{(3)}$. Instead, sum rules which involve one power of $\lambda_k$ or $b_{ijk}$ contribute to partially cancel parts of the amplitudes that scale slower than $s^3$ in $\mathcal{M}$.
 In the following, we will employ $q(y)=0$. In this case, the Liouville operator can be written as
\begin{equation}
\label{eq:L-product}
    L[f](y)\equiv -\dfrac{1}{r(y)}\dfrac{\diff}{\diff{y}}\left[p(y)\dfrac{\diff{f}}{\diff{y}}\right]\,.
\end{equation}
The linearity of the Liouville operator allows us to write the relation
\begin{equation}
    L[fg]=g L[f]+f L[g] -2\dfrac{p}{r}(\partial f)(\partial g)\,.
\end{equation}
For $\partial_y\psi_i|_{y=\pi}=0$ and by means of Eq.~\eqref{eq:linear-master-eq}-\eqref{eq:L-product}, one can show 
\begin{align}
        \label{eq:sum-rule-2}&\sum\limits_{k=0}^\infty\psi_k(\pi)\lambda_k a_{ijk}=\psi_i(\pi)\psi_j(\pi)(\lambda_i+\lambda_j)\,,\\
    &\sum\limits_{k=0}^\infty\psi_k(\pi)\lambda_k^2 a_{ijk}=\psi_i(\pi)\psi_j(\pi)(\lambda_i^2+6\lambda_i\lambda_j+\lambda_j^2)\,,
\end{align}
These sum rules generalize the ones in Ref.~\cite{deGiorgi:2020qlg, deGiorgi:2021xvm}.

 Similar sum rules can be obtained for the $b$-integrals. In principle, such relations are not strictly speaking needed, as one can use Eq.~\eqref{eq:b-from-a} and trade $b$- for $a$- coefficients. However, we report them here for completeness, as they save some work avoiding the conversion. For example,
\begin{equation}
    \begin{split}
        \sum\limits_{k=0}^\infty \psi_k(\pi)b_{ijk}&=  \sum\limits_{k=0}^\infty\psi_k(\pi) \int_\Omega \diff{y} \ p(y) \partial_y\psi_i\partial_y\psi_j \psi_k\\
        &= \int_\Omega \diff{y}\frac{p(y)}{r(y)}\partial_y\psi_i\partial_y\psi_j\sum\limits_{k=0}^\infty \ r(y)  \psi_k\psi_k(\pi) \\
        &= \int_\Omega \diff{y}\frac{p(y)}{r(y)}\partial_y\psi_i\partial_y\psi_j \delta(y-\pi)\\
        &=\left.\frac{p(y)}{r(y)}\partial_y\psi_i\partial_y\psi_j\right|_{y=\pi}  = 0\,,
    \end{split}
\end{equation}
where we have used the fact that $r(y)\neq 0$ and the boundary condition $\partial_y\psi_i(y)|_{y=\pi}=0$. 
Finally, through Eq.~\eqref{eq:general-sturm}, multiplying both sides by appropriate $\psi(\pi)$s and summing it over the appropriate indices, it can be  proved that
\begin{align}
    &\sum\limits_{k=0}^\infty \psi_k(\pi)b_{kij}=\lambda_i \psi_i(\pi)\psi_j(\pi) \,.
\end{align}
Explicit verification of these sum rules can be found in App.~\ref{app:LED} for LED and in Ref.~\cite{deGiorgi:2020qlg} for the large $\mu$ limit of RS. 
\end{subsection}
\end{section}
\begin{section}{Physics Applications}
\label{sec:physics_application}
The equation of motion for the $y$-component of the gravitons shown in Eq.~\eqref{eq:generic-metric-eq} can be matched to the SL equation with
    \begin{align}
        &\nonumber p(y)=e^{-3(l+1)k|y|}\,, &  q(y)=0  \,,\\ 
        &r(y)=e^{3(l-1)k|y|}\,, & \lambda _n=m_n^2 \,.
    \end{align}
Matching to the boundary conditions of the orbifold symmetry, the SL problem is defined by vanishing derivatives on the boundary, i.e.
\begin{equation}
    \partial\psi_i(0)=\partial\psi_i(\pi)=0\,,
\end{equation}
which makes it regular according to the definition of Eq.~\eqref{eq:regular-cond}.

The $a$- and $b$-coefficients enter the effective Lagrangian of the massive gravitons stemming from EH action, which can be decomposed as
\begin{equation}
    \begin{split}
        \mathcal{L}\supset&\, \sum\limits_{ij} \left(a_{ij}O_{ij}(h^2)+b_{ij}O'_{ij}(h^2)\right)\\
        &+\sum\limits_{ijk}\left(a_{ijk}O_{ijk}(h^3)+b_{ijk}O'_{ijk}(h^3)\right)+\dots\,,
    \end{split}
\end{equation}
where $O_{ij\dots}(h^n)$ is a an operator which involves $n$-powers of the field $h$ with indices $(i,j,\dots)$. The superscript~$'$ indicates that the operator originally contained two $y$-derivatives that have been absorbed in the definition of the corresponding $b$-coefficient (cfr.~Eq.~\eqref{eq:def-b}).
The normalization of the 5D wave functions discussed in the previous section is such that the $O^{(')}(h^2)$ part of the Lagrangian reproduces the Fierz-Pauli Lagrangian for a tower of massive gravitons~\cite{Fierz:1939ix}.
The explicit expansion up to $\mathcal{O}(h^4)$ can be found e.g. in Ref.s~\cite{Chivukula:2020hvi,deGiorgi:2020qlg}.
Regarding the matter content, we can consider a toy model with a scalar field such that
\begin{equation}
    S\supset \int_{y=\pi}\diff^4x\sqrt{-g}\left(\dfrac{1}{2}g^{\mu\nu}\partial_\mu\phi\partial_\nu\phi-\dfrac{1}{2}m_\phi^2\phi^2\right)\,.
\end{equation}
The conclusions do not depend on the matter content as long as it is localized on the brane.

Before moving forward, it is important to notice that LED, RS, and Clockwork share important features.
To make it more manifest, we move to conformal coordinates, $z$, defined such that the metric of Eq.~\eqref{eq:metric} becomes conformally flat
\begin{equation}
\label{eq:conformal}
    \diff{y} = A^{3l} \diff{z} \,,\qquad \diff s^2= A(z)^2\left(\eta_{\mu\nu}\diff{x}^\mu
\diff{x}^\nu -\diff z^2\right)\,,
\end{equation}
where
$A(y)\equiv e^{-k|y|} $.
In such coordinates, the SL equation is determined by the functions
\begin{equation}
\label{eq:conformal-2}
    r(z)=p(z)=A^3 \quad , \quad q(z)=0 \,.
\end{equation}
As formally there is no $l$-dependent term in these coordinates, it shows that all these models are conformally flat and equivalent, and hence characterized by the same structure of gravitons interactions in the 4D EFT. Furthermore, as the couplings stemming from integrating out the 5th dimension formally follow the same equations, they obey the same sum rules.

We now exemplify the applications in the RS model by demonstrating that the higher $s$ powers of the amplitude vanish after the sum rules are applied. Here we draw on the results of Ref.~\cite{deGiorgi:2020qlg}.
It turns out that the leading term in the amplitude is $\mathcal{M}^{(3)}$ and depends solely on graviton interactions.
Such amplitude is given by
    \begin{equation}
        \mathcal{M}^{(3)}=-\frac{i \sin ^2(\theta)}{24 M_5^3\lambda_i \lambda_j}\left(\sum\limits_{k=0}^\infty \psi_k(\pi)a_{ijk}-\psi_i(\pi)\psi_j(\pi)\right)\,.
    \end{equation}
    As can be seen, by means of Eq.~\eqref{eq:lin-0-sr} this contribution vanishes. Cancellation of the contributions to the amplitudes that scale slower than $s^3$ involves at least one power of $\lambda_k, b_{ijk}$ and/or the scalar mode of the metric, the radion. 
    The former sum rules are relevant to cancelling many
    terms of the amplitude both at $\mathcal{M}^{(2)}$ and $\mathcal{M}^{(3/2)}$ which are proportional to
    \begin{equation}
        \mathcal{M}^{(2),(3/2)}\propto\left(\sum\limits_{k=0}^\infty \psi_k(\pi)\lambda_k a_{ijk}-\psi_i(\pi)\psi_j(\pi)(\lambda_i+\lambda_j)\right)\,.
    \end{equation}
    They vanish due to Eq.~\eqref{eq:sum-rule-2}.
    Instead, the latter ones, which include the radion must be treated differently as the radion wave function and its normalization are not generated directly by the same SL equation of the gravitons. 
    As the treatment of the radion cannot be done in full generality as we did for the graviton, we show the amplitudes and the sum rules that cure them in the following section for the RS model.

\end{section}
\begin{section}{RS Radion Sum Rules beyond the large-$\mu$ limit}
\label{sec:radion-rules}
We supplement the linear sum rules presented the previous section with two additional sum rules that involve the radion  in the RS model. These are needed to cancel all contributions to the amplitude that grow faster than $s$, as shown for the scattering of KK-gravitons both in unstabilized and stabilized models~\cite{Chivukula:2022kju, Chivukula:2022tla}. The missing contribution corresponds to the diagram in Fig.~\ref{fig:radion-diagram}. We showed this previously in Ref.~\cite{deGiorgi:2020qlg} for the limiting case $\mu\equiv k R\gg 1$.  The proof was built on the explicit solutions of the SL equation in this limit.  The following discussion generalizes this to any value of $\mu$. Analogous results were published simultaneously to this paper in Ref.~\cite{Chivukula:2023sua}.

We work in the so-called \textit{unitary gauge}, i.e. in coordinates in which only the tensorial and scalar excitations of the metric are physical.
Similar techniques can be employed also for different theories. 

The couplings involving the radion are special as they do not depend on the SL functions, $\psi_i$. Their impact in the 4D effective couplings shows up by adding extra powers of $e^{-k|y|}$ in the n-points integrals of the gravitons. Even though this seems to eliminate the precious insights given to us by the SL theory, a geometric underlying connection inherited by the original 5D theory is still present. 
We will work in conformal coordinates, $z(y)$, as defined in Eq.~\eqref{eq:conformal}.
In such coordinates the difference in the definitions between a $a$ and $b$ coefficients is merely given by the presence of the two derivatives $\partial_z$ in the $b$-one, while they have the same powers of $A(z)$.

In conformal coordinates, for every radion in the expansion, a factor of $A^{-2}$ comes along. Therefore we define the couplings with the radion as
\begin{align}
        & \psi_r^2\IN{z}\, A(z)^{-1}\equiv1\,,\\
        &b_{ijr}\equiv \psi_r \IN{z}\, A(z) \partial_z \psi_i \partial_z \psi_j\ \,,\\
        &a_{irr} \equiv \psi_r^2 \IN{z}\, A(z)^{-1} \psi_i \,.
\end{align}
With a slight abuse of notation, we write $\psi_n(\pi)$ or $z=\pi$ in the integrals, which have to be understood as $z=\pi\sim z(y=\pi)$.
\begin{figure*}[tbh]
    \centering
     \subfigure[{}\label{fig:radion-diagram}]{\includegraphics[width=0.21\linewidth]{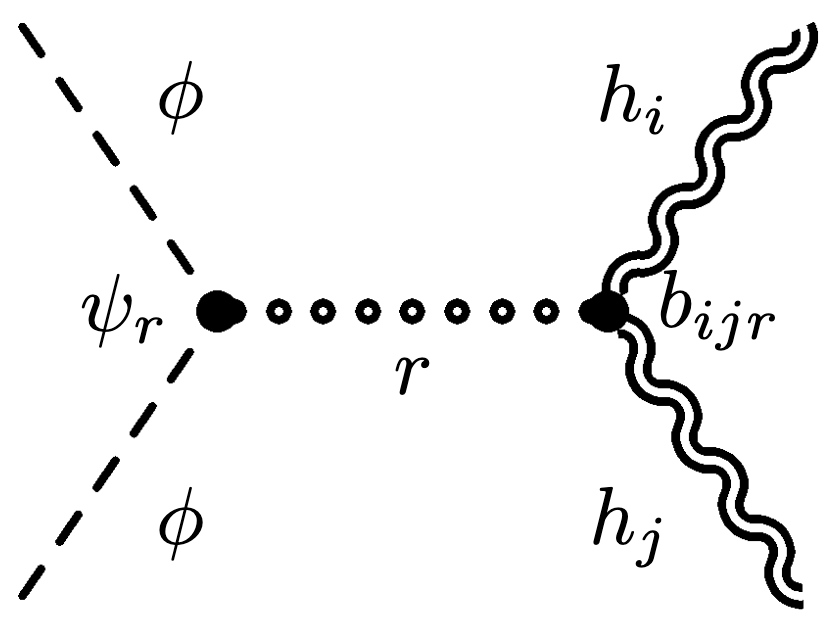}}
     \,\quad
     \subfigure[{}\label{fig:graviton-radion}]{\includegraphics[width=0.7\linewidth]{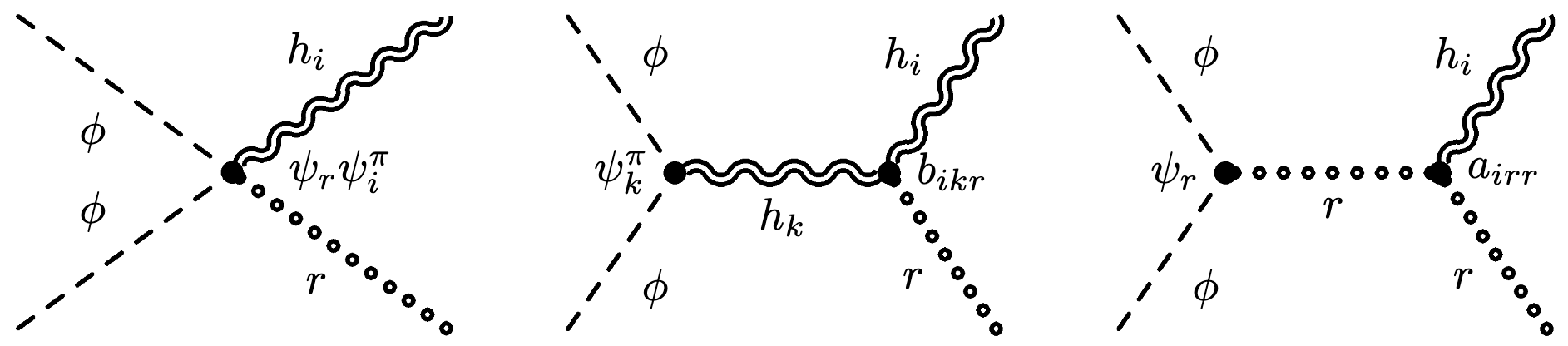}}
    \caption{Representative set of diagrams involving the radion for $\phi\phi\to h_i h_j$ (left) and $\phi\phi\to h_i r$ (right).} 
\label{fig:radion-diagrams}
\end{figure*}
    
    After imposing the graviton sum rules already derived, the non-vanishing amplitude at order $\mathcal{M}^{(2)}$ reads
\begin{widetext}
\begin{equation}
\begin{split}
    \mathcal{M}^{(2)}=-\frac{i s^2}{36 M_5^3 \lambda_i \lambda_j} & \left[\sum_{k=1}^\infty \dfrac{\psi_k(\pi)}{\lambda_k}a_{ijk}- \left(\lambda_i-\lambda_j\right)^2 (\lambda_i+\lambda_j)\left[\psi_i(\pi)\psi_j(\pi)+\psi_0^2\delta_{ij}\right]+6 \psi_rb_{ijr}\right]\,.
    \end{split}
    \label{eq:sum_s2}
\end{equation} 
\end{widetext}
At order $\mathcal{M}^{(3/2)}$ the amplitudes vanish employing the graviton sum rules, leaving the final amplitude of order $\mathcal{M}^{(1)}$. 
Finally, we consider matter annihilation into a massive graviton and a radion, $\phi\phi\to h_i r$. The corresponding diagrams are shown in Fig.~\ref{fig:graviton-radion}. The leading amplitude is of order $\mathcal{M}^{(2)}$ and reads
\begin{equation}
\mathcal{M}^{(2)}=\dfrac{i s^2}{24 M_5^3 \lambda_i} \left(\sum\limits_{k=1}^\infty\psi_k(\pi)\frac{b_{ikr}}{\lambda_k}+\psi_r a_{irr}-A_\pi^{-2}\psi_r\psi_i(\pi)\right)\,,
\end{equation}
To cancel these contributions two additional sum rules are required.

The starting point for their proof is the identity
\begin{equation}
\label{eq:fun2}
    A(\pi)^{\alpha-3}f(\pi)=\sum\limits_{n=0}^\infty\psi_n(\pi)\left(\IN{z}A(z)^\alpha \psi_n(z)f(z)\right)\,.
\end{equation}
Through the SL equation and integrating by parts, we can write the above expression as
\begin{equation}
        \begin{split}
            &A(\pi)^{\alpha-3}f(\pi)=\psi_0^2\left(\IN{z}A(z)^\alpha f(z)\right)\\
            &+\sum\limits_{n=1}^\infty\dfrac{\psi_n(\pi)}{\lambda_n}\left(\IN{z}A(z)^3 \partial\psi_n(z)\partial\left(A^{\alpha-3}f(z)\right)\right)\,.
        \end{split}
\end{equation}
Now we choose
\begin{equation}
    f(z)=A(z)^{3-\alpha}\int\limits_\pi^z\diff{z'}A^{\beta-3}g(z')\,,
\end{equation}
such that we obtain the result of interest:
\begin{equation}
\label{eq:fundmanetal-2}
     \begin{split}
         &\sum\limits_{n=1}^\infty\dfrac{\psi_n(\pi)}{\lambda_n}\left(\IN{z}A(z)^\beta \partial\psi_n(z)g(z)\right)\\
         &=-\psi_0^2\left(\IN{z}A(z)^3 \int\limits_\pi^z\diff{z'}A^{\beta-3}g(z')\right)\,.
     \end{split}
\end{equation}
This will be a key ingredient to proving the two sum rules. Its convenience relies upon the possibility of trading the infinite sum with the eigenvalues in the denominator into a double integral.

\paragraph{Sum Rule 1.}
We are interested in computing 
\begin{equation}
    \sum\limits_{k=1}^\infty \dfrac{\psi_k(\pi)}{\lambda_k}a_{ijk}(\lambda_i-\lambda_j)^2\,.
\end{equation}
By means of Eq.~\eqref{eq:fundmanetal-2}, it is convenient to define the following quantity
\begin{equation}
    S_{ij}\equiv\sum\limits_{k=1}^\infty\dfrac{\psi_k(\pi)}{\lambda_k}b_{kij}=-\psi_0^2\IN{z}A^3\int\limits_\pi^z\partial\psi_i\psi_j\,.
\end{equation}
Via Eq.~\eqref{eq:b-from-a}, one proves that the problem is then reduced to
\begin{equation}
    \sum\limits_{k=1}^\infty \dfrac{\psi_k(\pi)}{\lambda_k}a_{ijk}(\lambda_i-\lambda_j)^2=(S_{ij}-S_{ji})(\lambda_i-\lambda_j)\,.
\end{equation}
The strategy is to prove two independent relations for $S_{ij}$ and to solve for the latter.
The first relation can be obtained pretty straightforwardly
\begin{equation}
\label{eq:rel1}
    S_{ij}+S_{ji}=\psi_i(\pi)\psi_j(\pi)-\psi_0^2\delta_{ij}\,.
\end{equation}
The second one requires some work. Similarly to the approach used in Ref.~\cite{deGiorgi:2020qlg}, we consider
\begin{equation}
    \begin{split}
        &\lambda_jS_{ij}+\lambda_i S_{ji}\\
        &=\psi_0^2\left[\lambda_i\delta_{ij}+6\IN{z}A^3\int\limits_\pi^z\diff{z'}A^{-1}\partial(A)\partial\psi_i\partial\psi_j\right]
    \end{split}
\end{equation}
Given that $\diff{y}\equiv A\diff{z}$, it follows that~\footnote{This is the point where the value of $l$ would have entered the discussion if we kept it unspecified.}
\begin{equation}
    \partial_z(A)=A \partial_y(e^{-\mu y})=-\mu A^{2}\,.
\end{equation}
Let us denote for simplicity $A_\pi\equiv A(\pi)$ and consider
\begin{widetext}
\begin{equation}
\label{eq:example-good1}
    \begin{split}
        \IN{z}A^\beta\partial\psi_i\partial\psi_j&=\IN{z}A^{\alpha+\beta}\partial\int\limits_\pi^zA^{-\alpha}\partial\psi_i\partial\psi_j=A_\pi^{\alpha+\beta}\IN{z}A^{-\alpha}\partial\psi_i\partial\psi_j-(\alpha+\beta)\IN{z}A^{\alpha+\beta-1}\partial(A)\int\limits_\pi^zA^{-\alpha}\partial\psi_i\partial\psi_j\\
        &=A_\pi^{\alpha+\beta}\IN{z}A^{-\alpha}\partial\psi_i\partial\psi_j+\mu(\alpha+\beta)\IN{z}A^{\alpha+\beta+1}\int\limits_\pi^zA^{-\alpha}\partial\psi_i\partial\psi_j\,,
    \end{split}
\end{equation}
\end{widetext}
where we have also used that $A(-\pi)=A(\pi).$
We choose $\alpha=-1$ and $\beta=3$ to find
\begin{equation}
    \begin{split}
        &\lambda_i \delta_{ij}=A_\pi^{2}\IN{z}A\partial\psi_i\partial\psi_j+2\mu\IN{z}A^{3}\int\limits_\pi^zA\partial\psi_i\partial\psi_j\,,\\
        &\,=A_\pi^{2}\IN{z}A\partial\psi_i\partial\psi_j-2\IN{z}A^{3}\int\limits_\pi^zA^{-1}\partial(A)\partial\psi_i\partial\psi_j\,.
    \end{split}
\end{equation}
We shuffle the terms to write the main result of interest
\begin{equation}
    \begin{split}
        &2\IN{z}A^{3}\int\limits_\pi^zA^{-1}\partial(A)\partial\psi_i\partial\psi_j\\
        &=-\lambda_i \delta_{ij}+A_\pi^{2}\IN{z}A\partial\psi_i\partial\psi_j\,.
    \end{split}
\end{equation}
We recognize the last integral term to be $\psi_r^{-1}b_{ijr}$.
This implies that 
\begin{equation}
\label{eq:rel2}
    \lambda_jS_{ij}+\lambda_i S_{ji}=\psi_0^2\left\{3 A_\pi^{2}\psi_r^{-1}b_{ijr}-2\lambda_i\delta_{ij}\right\}\,.
\end{equation}
By combining Eq.s~\eqref{eq:rel1}-\eqref{eq:rel2}, we get the desired result
\begin{equation}
    \begin{split}
        &S_{ij}=\dfrac{\lambda_i\left[\psi_i(\pi)\psi_j(\pi)+\psi_0^2\delta_{ij}\right]-3A_\pi^{2}\psi_r^{-1}\psi_0^2 b_{ijr}}{\lambda_i-\lambda_j}\,,
    \end{split}
\end{equation}
which finally leads to
\begin{equation}
    \begin{split}
        &\sum\limits_{k=1}^\infty \dfrac{\psi_k(\pi)}{\lambda_k}a_{ijk}(\lambda_i-\lambda_j)^2=(S_{ij}-S_{ji})(\lambda_i-\lambda_j)\,,\\
        &\,=(\lambda_i+\lambda_j)\left[\psi_i(\pi)\psi_j(\pi)+\psi_0^2\delta_{ij}\right]-6 A_\pi^2\psi_r^{-1}\psi_0^2 b_{ijr}\,.
    \end{split}
\end{equation}
The presence of $\psi_0^2$ is at first glance somewhat surprising as the massless graviton does not necessarily play a role in the amplitude, e.g. in the limit $\mu\gg 1$. We can replace it by noticing that
\begin{equation}
    \psi_0^2 = \frac{\mu}{1-e^{-2\mu\pi}} \quad , \quad \psi_r^2 = \frac{\mu }{e^{2 \pi  \mu }-1}\,,
\end{equation}
so that
\begin{equation}
    \label{eq:norms}
    \psi_0^2 A_\pi^2 = \psi_r^2\,.
\end{equation}
Thus we have
\begin{equation}
    \begin{split}
        &\sum\limits_{k=1}^\infty \dfrac{\psi_k(\pi)}{\lambda_k}a_{ijk}(\lambda_i-\lambda_j)^2\\
        &=(\lambda_i+\lambda_j)\left[\psi_i(\pi)\psi_j(\pi)+\psi_0^2\delta_{ij}\right]-6 \psi_r b_{ijr}\,.
    \end{split}
\end{equation}
This proves the sum rule.

\paragraph{Sum Rule 2.} 

We use Eq.~\eqref{eq:fundmanetal-2} to find
\begin{equation}
    \begin{split}
        &\sum\limits_{k=1}^\infty\psi_k(\pi) \dfrac{b_{kir}}{\lambda_k}= -\psi_0^2\psi_r\IN{z}A(z)^3\int\limits_\pi^z A^{-2}\partial\psi_i\,,\\
        &\,=-\psi_0^2\psi_r\left[c_i-\psi_0^{-2}A_\pi^{-2}\psi_i(\pi)-2\mu\IN{z}A^3\int\limits_\pi^z\psi_i A^{-1}\right]\,,
    \end{split}
\end{equation}
where $c_i\equiv\IN{z}A\psi_i$.

Using the same strategy used in Eq.~\eqref{eq:example-good1}, one can prove that
\begin{equation}
    \begin{split}
        &\mu(\alpha+\beta)\IN{z}A^{\alpha+\beta+1}\int\limits_\pi^z A^{-\alpha}\psi_i\\
        &=\IN{z}A^\beta\psi_i-A_\pi^{\alpha+\beta}\IN{z}A^{-\alpha}\psi_i\,.
    \end{split}
\end{equation}
We can now match the previous result by choosing $\beta=1$ and $\alpha=1$ such that $\alpha+\beta=2$. Remarkably, by merging the two contributions, the $c_i$ part cancels and leads to the final result
\begin{equation}
    \begin{split}
        \sum\limits_{k=1}^\infty\psi_k(\pi) \dfrac{b_{kir}}{\lambda_k}&=A_\pi^{-2}\psi_r\psi_i(\pi)-\psi_0^{2}\psi_r^{-1}A_\pi^{2}a_{irr}\\
        &=A_\pi^{-2}\psi_r\psi_i(\pi)-\psi_r a_{irr}\,.
    \end{split}
\end{equation}
This concludes the proof.
\end{section}
\begin{section}{Summary and Outlook}
\label{sec:summary_and_outlook}
In this work, we have proven in full generality a set of sum rules for solutions to the Sturm-Lioville~(SL) problem. This generalizes our previous results put forward in \cite{deGiorgi:2020qlg} where similar conclusions were reached for the large $\mu$ limit of the RS model.  We have then applied them and shown their power in the framework of orbifolded extra-dimensional gravity, with particular emphasis on LED, RS, and CW models. We exploit the fact that these models are conformally equivalent and thus formally possess the same KK-graviton interactions.
In graviton pair-productions from matter, the sum rules have been shown to cancel the unphysical high-energy growth of the amplitudes, reducing them from  $\mathcal{O}(s^3)\to \mathcal{O}(s^2)$. 
The full reduction of the amplitudes to $\mathcal{O}(s)$ requires the contribution of the radion.
These depend on the compactification and we have not found a general form. However, we give an example of how they can be found in the RS-model without making simplifying assumption regarding the wave-functions.

Regarding the amplitudes that do not involve the radion, we would like to comment on the applicability of our results to other models in the following. The 5D theory and its 4D limit need to fulfil a number of conditions for our approach to work. We need the 4D theory to consist of spin-2 fields with a Fierz-Pauli action. The 5D theory has to possess a covariant action with at most two derivatives and matter is localized on a brane. Under these conditions:
\begin{enumerate}
    \item Regardless of cancellations in the amplitude, the sum rules still apply. If the action contains at most two derivatives, then the 5D part of each graviton will satisfy a second-order linear differential equation. All second-order linear differential equations can be recast as SL equations by multiplying them by an appropriate integration factor. The diagrams that can contribute to the scattering processes considered do not change, so the sum rules that can enter are only those derived in the previous section.
\item  In such theories, the action fixes
\begin{equation}
    q(y)=0\,.
\end{equation}
This is an important requirement for the derivation of some of the sum rules, which should be otherwise slightly modified. It can be understood by considering the derivative structure of the action. As it can contain at most two 5D derivatives of the fields, in the basis in which the 5D graviton does not mix with any other component of the metric, only terms with two purely $\partial_\mu$ or $\partial_5$ can be present. Therefore, a term of the $q(y)$-type would necessarily include two $\partial_\mu$, which belong to the kinetic term part of the Lagrangian, not to the mass part. Furthermore, from a physical point of view, this condition guarantees the presence of a massless KK-graviton (cfr. Eq.~\eqref{eq:SL}), since it ensures that a constant wave function with a zero eigenvalue is a solution.
\item Using the previous argument, the corresponding SL equations are also expected to have $q(y)=0$. The possible relevant differences with respect to what is derived in this work depend on the type of compactification and they are hidden in the boundary conditions. In the case of orbifold compactification, we used
\begin{equation}
    \left.\dfrac{\diff\psi_i}{\diff{y}}\right|_{\partial\Omega}=0\,.
\end{equation}
The absence of such a condition might result in a modification of some results, which has then to be taken into account during the derivation of the sum rules.
\end{enumerate}

 The cancellations of $\mathcal{M}^{(n)}$ for $n>1$ that are encoded in the sum rules have a very notable impact in phenomenological studies involving KK-graviton production, e.g. in studies of the early universe where high-momentum transfer is common. Such cancellations are possible thanks to the underlying geometric structure of the original 5D theory. 
\end{section}

\section*{Acknowledgments} 
A.d.G. thanks Alejandro Pérez Rodríguez for useful discussions and especially thanks Carlos A. Argüelles and the group of Palfrey House for their hospitality and the stimulating working environment during which a core part of this work was realised.
The work of A.d.G. is supported by the European Union's Horizon 2020 Marie Sk\l odowska-Curie grant agreement No 860881-HIDDeN.

\section*{Note added}

The initial version of this work contained an extensive discussion of the sum rules for graviton-scattering. After the first version appeared on the arXiv, it was brought to our attention that the proof of the sum rules reported in Ref.~\cite{Chivukula:2020hvi} is also based on the properties of solutions of the Sturm-Liouville problem and not limited to the Randall-Sundrum model. Therefore, this proof is already general and we decided to remove this part from the updated version of this work.

\begin{appendices}
\numberwithin{equation}{section}
\section{Explicit Example: Large Extra-Dimensions}
\label{app:LED}

\subsection{The Model}
The LED~\cite{Arkani-Hamed:1998jmv} can be considered as the RS in the limit $k \to 0$ (thus $\mu\to 0$) while $R$ is kept fixed. As in the main text, we will work in units of $R$, or, equivalently, we will set $R=1$.
The numerical coefficients that enter the sum rules in Table~\ref{eq:sumrules} can be computed exactly.
The EOMs of $\psi_n$ are given by
\begin{equation}
    \left(\partial_y^2+m_n^2\right)\psi_n=0 \,,
\end{equation}
with solutions
\begin{align}
    &\psi_0(y) = N_0\left[1+\alpha_0 |y|\right] \,, \\
    &\psi_n(y)=N_n \left[\cos{\left(m_n |y|\right)}+\alpha_n \sin{\left(m_n |y|\right)}\right] \,,
\end{align}
where we have set $m_0=0$.
Imposing the boundary conditions $\partial_y \psi_n=0$, we need to set $\alpha_n=\alpha_0 = 0$, and hence
\begin{equation}
    \psi_0 = N_0 \quad , \quad \psi_n(y)=N_n\cos{\left(m_n y\right)} \,.
\end{equation}
By periodicity of the solutions, we can also fix the masses
\begin{equation}
    \cos{(m_n y)} = \cos{(m_n (y+2\pi))} \quad \Rightarrow \quad m_n =n  \,.
\end{equation}
With such definitions the wave functions are orthogonal and normalized~\footnote{We have chosen here to incorporate in the scalar product an overall factor of $1/\pi$.}
\begin{equation}
    \frac{1}{\pi}\int\limits_{-\pi}^\pi \diff{y}\ \psi_i(y)\psi_j(y) =  \delta_{ij} \,,
\end{equation}
fixing
\begin{equation}
   N_0 = \frac{1}{\sqrt{2}} \quad , \quad N_n = 1 \,.
\end{equation}
To simplify the discussion and avoid redundant $\pm$ signs, we consider the matter content to be localized on the brane at $y=0$. This does not affect any of the results expected from the discussion in the main text.
Thus, the fifth-dimensional wave functions are given by
\begin{equation}
    \begin{split}
        &\psi_0(y)= N_0=\frac{1}{\sqrt{2}}\\
        &\psi_n(y)=N_n\cos{(ny)}=\cos{(ny)} \,.
    \end{split}
\end{equation}

\subsection{Sum Rules}
We turn now to the sum rules. 
According to the definitions of Eq.s~\eqref{eq:def-a} and \eqref{eq:def-b}, we define
\begin{align}
    &a_{ijk}\equiv N_i N_j N_k\, \chi_{ijk}\,,\\
    &b_{ijk}\equiv N_i N_j N_k\, \tilde{\chi}_{ijk} \,, 
\end{align}
so that 
\begin{align}
    & \chi_{ijk}\equiv \frac{1}{\pi}\int\limits_{-\pi}^\pi \diff{y}\ \cos(iy)\cos(jy)\cos(ky) \,, \label{chi}\\
    & \tilde{\chi}_{ijk}\equiv  \frac{ij}{\pi}\int\limits_{-\pi}^\pi \diff{y}\ \sin(iy)\sin(jy)\cos(ky) \label{chi-tilde}\,.
\end{align}
One can calculate explicitly the integrals which gives the rather simple expressions 
\begin{align}
\label{eq:chi}
    &\chi_{ijk}=\frac{1}{2}\left[\delta_{(i-j-k),0}+\delta_{(i+j-k),0}+\delta_{(i-j+k),0}+\delta_{(i+j+k),0}\right] \,,  \\
    &\tilde{\chi}_{ijk}=\frac{ij}{2}\left[\delta_{(i-j-k),0}-\delta_{(i+j-k),0}+\delta_{(i-j+k),0}-\delta_{(i+j+k),0}\right] \label{eq:chi-tilde} \,,
\end{align}
where $\delta_{i,j}$ is the Kronecker delta.
The above quantities satisfy Eq.~\eqref{eq:general-sturm},
\begin{equation}
\label{eq:general-sturm-chi}
    i^2 \chi_{ijk} = \tilde{\chi}_{ijk}+\tilde{\chi}_{ikj}\,.
\end{equation}
Similarly, we define the coefficients associated with the radion as
\begin{align}
   &\tilde{b}_{ijr}\to  N_i N_j N_r\,\tilde{\chi}_{ijr}\,,\\
   &a_{irr} \to N_i N_r^2\, \chi_{irr}\,,
\end{align}
with
\begin{align}
    &\tilde{\chi}_{ijr}\equiv \frac{ij}{\pi}\int\limits_{-\pi}^\pi \diff{y}\sin(iy)\sin(jy)=i^2\delta_{i,j}\,,\label{chir} \\
    &\chi_{irr}\equiv \frac{1}{\pi}\int\limits_{-\pi}^\pi \diff{y} \cos(iy) = 2\delta_{i,0}\,.
\end{align}
The sum rules needed for the cancellation of the gravitons pair-production from matter amplitudes, counterparts of those in Table~\ref{eq:sumrules}, are given in Table~\ref{tab:sum_rules}. 
\begin{table}[tbh]
\centering
\begin{tabular}{cc}
\toprule
SR 1: & $\sum\limits_{k=0}^\infty N_k^2 \chi_{ijk}=1$\\
SR 2: & $\sum\limits_{k=0}^\infty N_k^2 k^2\chi_{ijk}=i^2+j^2$ \\
SR 3: & $\sum\limits_{k=0}^\infty N_k^2 \tilde\chi_{kij}=i^2$\\
SR 4: & $\sum\limits_{k=0}^\infty N_k^2 \tilde\chi_{ijk}=0$\\
SR 5: &$\sum\limits_{k=1}^\infty N_k^2 N_i N_j \frac{\chi_{ijk}}{k^2}(i^2-j^2)^2=\left(i^2+j^2\right)\left(N_i N_j +N_0^2\delta_{ij}\right)$\\
& $-6 N_r^2\tilde{\chi}_{ijr}$\\
SR 6:
& $\sum\limits_{j=1}^\infty N_j^2\frac{\tilde{\chi}_{ijr}}{j^2}=1-N_r^2 \chi_{irr}$\\
\bottomrule
\end{tabular}
\caption{Sum Rules (SR) needed for the cancellations.\label{tab:sum_rules}}
\label{eq:allsumrules}
\end{table}

\paragraph{Sum Rule 1} Let us focus first on $\chi_{ijk}$ proving that
\begin{equation}
\label{sum-1}
    \sum\limits_{k=0}^\infty N_k^2 \chi_{ijk}=1 \ .
\end{equation}
It is clear from the definition of Eq.~\eqref{eq:chi} that $\chi_{00k}=2\delta_{k,0}$ thus the sum rule is satisfied for $i=j=0$ as $N_0^2=1/2$. If $i\neq j = 0$, then $\chi_{i0k}=\delta_{ki}$ which sets the sum to $1$, as $N_{i\neq 0}=1$. Finally, in case $i,j\neq 0$, one can easily check that just two terms contribute with $1/2$, thus giving $1/2+1/2=1$, proving the last case of the sum rule.
\paragraph{Sum Rules 2,3,4} Let us now focus on the sum rules involving $k^2 \chi_{ijk}$. First, we look at $\sum\limits_{k=0}^\infty N_k^2 \tilde{\chi}_{ijk}$. It is clear from Eq.~\eqref{eq:chi-tilde} that $\tilde{\chi}_{0jk}=\tilde{\chi}_{i0k}=\tilde{\chi}_{00k}=0$, thus the only non trivial case is for $i,j\neq 0$. However, as $k$ runs from $0$ to $\infty$, it is clear that only two terms contribute and they are opposite in sign, i.e. $\sum\limits_{k=0}^\infty N_k^2 \tilde{\chi}_{ijk} \propto (1-1)=0$. In general, then we have that
\begin{equation}
\label{sum-2}
    \sum\limits_{k=0}^\infty N_k^2 \tilde{\chi}_{ijk}  = 0 \ .
\end{equation}
The other rules can be obtained using the sum rules we have derived up to now with Eq.~\eqref{eq:general-sturm-chi}, which implies
\begin{equation}
      N_j^2 i^2 \chi_{ijk} =  N_j^2 (\tilde{\chi}_{ijk}+\tilde{\chi}_{ikj}) \ .
\end{equation}
Summing over the index-$j$ and using both Eq.s~\eqref{sum-1} and \eqref{sum-2} we get
\begin{equation}
   \sum\limits_{j=0}^\infty N_j^2 \tilde{\chi}_{ijk} = i^2 \ .
\end{equation}
Finally, we can use Eq.~\eqref{eq:general-sturm-chi} and the above relations again to find the second sum rule of interest
\begin{equation*}
    \sum\limits_{k=0}^\infty N_k^2 k^2 \chi_{ijk} = i^2+j^2 \ .
\end{equation*}
It is instructive to make an explicit check of such relation since we have not verified all of them explicitly; let us assume without loss of generality that $j>k\neq 0$, then
\begin{equation}
  \begin{split}
      \sum\limits_{k=0}^\infty k^2 \chi_{kij} &=\frac{1}{2}\left[(i+j)^2+(i-j)^2\right]\\
      &=\frac{1}{2}\left[2i^2+2j^2\right]= i^2+j^2 \,,
  \end{split}
\end{equation}
which is what we expected.
\paragraph{Sum Rule 5} We need to prove that $\sum\limits_{k=1}^\infty N_k^2 \frac{\chi_{ijk}}{k^2}(i^2-j^2)^2=\left(i^2+j^2\right)\left(1+N_0^2\chi_{ij0}\right)-6N_r^2\tilde{\chi}_{ijr}$. The fastest method is to verify it explicitly. Looking at the definitions of the coefficients of Eq.s~\eqref{chi} and \eqref{chir}, it is clear that if $i=j=0$ the sum rule is trivially satisfied. The right-hand-side of the sum rule can be written as $\left(i^2+j^2\right)\left(1+1/2\delta_{ij}\right)-3i^2\delta_{i,j}$, so let us consider the case $i=j\neq0$ and $i\neq j$ separately. In the first case, the left-hand-side is $0$, while the right-hand-side becomes $2i^2\cdot 3/2-3i^2 = 0$, and thus it is satisfied. The case in which $i\neq j=0$ can be easily verified too as most of the terms disappear. Finally if $i\neq j\neq0$
\begin{equation}
   \begin{split}
       \sum\limits_{k=1}^\infty N_k^2 \frac{\chi_{ijk}}{k^2}(i^2-j^2)^2 &= \frac{(i^2-j^2)^2}{2}\left[\frac{1}{(i+j)^2}+\frac{1}{(i-j)^2}\right]\,,\\
       &=(i^2-j^2)^2\frac{1}{2}\frac{2(i^2+j^2)}{(i^2-j^2)^2}  = i^2+j^2 \ ,
   \end{split}
\end{equation}
which proves the last scenario and thus the sum rule.
\paragraph{Sum Rule 6}
The sum rule we are trying to prove is $\sum\limits_{j=1}^\infty N_j^2 \frac{\tilde{\chi}_{ijr}}{j^2}=1-N_r^2 \chi_{irr}=1-\delta_{i,0}$. In the case $i=0$ the proof is trivial as $\tilde{\chi}_{0jr}=0$ and hence we get $0=0$. If $i\neq 0$ we need to check that  $\sum\limits_{j=1}^\infty N_j^2 \frac{\tilde{\chi}_{ijr}}{j^2}=1$. This holds as  $\sum\limits_{j=1}^\infty N_j^2 \frac{\tilde{\chi}_{ijr}}{j^2}=\sum\limits_{j=1}^\infty \frac{i}{j}\delta_{i,j}=1$, concluding the proof.

$\left.\right.$

\end{appendices}

\bibliographystyle{hieeetr.bst}
\bibliography{Bibliography_Draft.bib}

\providecommand{\href}[2]{#2}\begingroup\raggedright\begin{thebibliography}{10}

\bibitem{Fierz:1939ix}
M.~Fierz and W.~Pauli, {\it {On relativistic wave equations for particles of
  arbitrary spin in an electromagnetic field}},  Proc. Roy. Soc. Lond. A {\bf
  173} (1939) 211--232.

\bibitem{Arkani-Hamed:2002bjr}
N.~Arkani-Hamed, H.~Georgi, and M.~D. Schwartz, {\it {Effective field theory
  for massive gravitons and gravity in theory space}},  Annals Phys. {\bf 305}
  (2003) 96--118, [\href{http://arxiv.org/abs/hep-th/0210184}{{\tt
  hep-th/0210184}}].

\bibitem{deRham:2010kj}
C.~de~Rham, G.~Gabadadze, and A.~J. Tolley, {\it {Resummation of Massive
  Gravity}},  Phys. Rev. Lett. {\bf 106} (2011) 231101,
  [\href{http://arxiv.org/abs/1011.1232}{{\tt arXiv:1011.1232}}].

\bibitem{Hinterbichler:2011tt}
K.~Hinterbichler, {\it {Theoretical Aspects of Massive Gravity}},  Rev. Mod.
  Phys. {\bf 84} (2012) 671--710, [\href{http://arxiv.org/abs/1105.3735}{{\tt
  arXiv:1105.3735}}].

\bibitem{Schwartz:2003vj}
M.~D. Schwartz, {\it {Constructing gravitational dimensions}},  Phys. Rev. D
  {\bf 68} (2003) 024029, [\href{http://arxiv.org/abs/hep-th/0303114}{{\tt
  hep-th/0303114}}].

\bibitem{SekharChivukula:2019yul}
R.~Sekhar~Chivukula, D.~Foren, K.~A. Mohan, D.~Sengupta, and E.~H. Simmons,
  {\it {Scattering amplitudes of massive spin-2 Kaluza-Klein states grow only
  as ${\cal O}(s)$}},  Phys. Rev. D {\bf 101} (2020), no.~5 055013,
  [\href{http://arxiv.org/abs/1906.11098}{{\tt arXiv:1906.11098}}].

\bibitem{SekharChivukula:2019qih}
R.~Sekhar~Chivukula, D.~Foren, K.~A. Mohan, D.~Sengupta, and E.~H. Simmons,
  {\it {Sum Rules for Massive Spin-2 Kaluza-Klein Elastic Scattering
  Amplitudes}},  Phys. Rev. D {\bf 100} (2019), no.~11 115033,
  [\href{http://arxiv.org/abs/1910.06159}{{\tt arXiv:1910.06159}}].

\bibitem{Chivukula:2020hvi}
R.~S. Chivukula, D.~Foren, K.~A. Mohan, D.~Sengupta, and E.~H. Simmons, {\it
  {Massive Spin-2 Scattering Amplitudes in Extra-Dimensional Theories}},  Phys.
  Rev. D {\bf 101} (2020), no.~7 075013,
  [\href{http://arxiv.org/abs/2002.12458}{{\tt arXiv:2002.12458}}].

\bibitem{Bonifacio:2019ioc}
J.~Bonifacio and K.~Hinterbichler, {\it {Unitarization from Geometry}},  JHEP
  {\bf 12} (2019) 165, [\href{http://arxiv.org/abs/1910.04767}{{\tt
  arXiv:1910.04767}}].

\bibitem{deGiorgi:2020qlg}
A.~de~Giorgi and S.~Vogl, {\it {Unitarity in KK-graviton production: A case
  study in warped extra-dimensions}},  JHEP {\bf 04} (2021) 143,
  [\href{http://arxiv.org/abs/2012.09672}{{\tt arXiv:2012.09672}}].

\bibitem{Chivukula:2022tla}
R.~S. Chivukula, D.~Foren, K.~A. Mohan, D.~Sengupta, and E.~H. Simmons, {\it
  {Spin-2 Kaluza-Klein scattering in a stabilized warped background}},  Phys.
  Rev. D {\bf 107} (2023), no.~3 035015,
  [\href{http://arxiv.org/abs/2206.10628}{{\tt arXiv:2206.10628}}].

\bibitem{Chivukula:2023sua}
R.~S. Chivukula, J.~A. Gill, K.~A. Mohan, D.~Sengupta, E.~H. Simmons, and
  X.~Wang, {\it {Scattering Amplitudes of Massive Spin-2 Kaluza-Klein States
  with Matter}},  \href{http://arxiv.org/abs/2311.00770}{{\tt
  arXiv:2311.00770}}.

\bibitem{Lee:2013bua}
H.~M. Lee, M.~Park, and V.~Sanz, {\it {Gravity-mediated (or Composite) Dark
  Matter}},  Eur. Phys. J. C {\bf 74} (2014) 2715,
  [\href{http://arxiv.org/abs/1306.4107}{{\tt arXiv:1306.4107}}].

\bibitem{Rueter:2017nbk}
T.~D. Rueter, T.~G. Rizzo, and J.~L. Hewett, {\it {Gravity-Mediated Dark Matter
  Annihilation in the Randall-Sundrum Model}},  JHEP {\bf 10} (2017) 094,
  [\href{http://arxiv.org/abs/1706.07540}{{\tt arXiv:1706.07540}}].

\bibitem{Babichev:2016bxi}
E.~Babichev, L.~Marzola, M.~Raidal, A.~Schmidt-May, F.~Urban, H.~Veerm\"ae, and
  M.~von Strauss, {\it {Heavy spin-2 Dark Matter}},  JCAP {\bf 09} (2016) 016,
  [\href{http://arxiv.org/abs/1607.03497}{{\tt arXiv:1607.03497}}].

\bibitem{Kraml:2017atm}
S.~Kraml, U.~Laa, K.~Mawatari, and K.~Yamashita, {\it {Simplified dark matter
  models with a spin-2 mediator at the LHC}},  Eur. Phys. J. C {\bf 77} (2017),
  no.~5 326, [\href{http://arxiv.org/abs/1701.07008}{{\tt arXiv:1701.07008}}].

\bibitem{Folgado:2019sgz}
M.~G. Folgado, A.~Donini, and N.~Rius, {\it {Gravity-mediated Scalar Dark
  Matter in Warped Extra-Dimensions}},
  \href{http://arxiv.org/abs/1907.04340}{{\tt arXiv:1907.04340}}. [Erratum:
  JHEP 02, 129 (2022)].

\bibitem{deGiorgi:2021xvm}
A.~de~Giorgi and S.~Vogl, {\it {Dark matter interacting via a massive spin-2
  mediator in warped extra-dimensions}},  JHEP {\bf 11} (2021) 036,
  [\href{http://arxiv.org/abs/2105.06794}{{\tt arXiv:2105.06794}}].

\bibitem{deGiorgi:2022yha}
A.~de~Giorgi and S.~Vogl, {\it {Warm dark matter from a gravitational freeze-in
  in extra dimensions}},  JHEP {\bf 04} (2023) 032,
  [\href{http://arxiv.org/abs/2208.03153}{{\tt arXiv:2208.03153}}].

\bibitem{Gill:2023kyz}
J.~A. Gill, D.~Sengupta, and A.~G. Williams, {\it {Graviton-photon production
  with a massive spin-2 particle}},  Phys. Rev. D {\bf 108} (2023), no.~5
  L051702, [\href{http://arxiv.org/abs/2303.04329}{{\tt arXiv:2303.04329}}].

\bibitem{Gonzalo:2022jac}
E.~Gonzalo, M.~Montero, G.~Obied, and C.~Vafa, {\it {Dark Dimension Gravitons
  as Dark Matter}},  \href{http://arxiv.org/abs/2209.09249}{{\tt
  arXiv:2209.09249}}.

\bibitem{Anchordoqui:2022svl}
L.~A. Anchordoqui, I.~Antoniadis, and D.~Lust, {\it {Aspects of the dark
  dimension in cosmology}},  Phys. Rev. D {\bf 107} (2023), no.~8 083530,
  [\href{http://arxiv.org/abs/2212.08527}{{\tt arXiv:2212.08527}}].

\bibitem{Giudice:2016yja}
G.~F. Giudice and M.~McCullough, {\it {A Clockwork Theory}},  JHEP {\bf 02}
  (2017) 036, [\href{http://arxiv.org/abs/1610.07962}{{\tt arXiv:1610.07962}}].

\bibitem{Arkani-Hamed:1998jmv}
N.~Arkani-Hamed, S.~Dimopoulos, and G.~R. Dvali, {\it {The Hierarchy problem
  and new dimensions at a millimeter}},  Phys. Lett. B {\bf 429} (1998)
  263--272, [\href{http://arxiv.org/abs/hep-ph/9803315}{{\tt hep-ph/9803315}}].

\bibitem{Randall:1999ee}
L.~Randall and R.~Sundrum, {\it {A Large mass hierarchy from a small extra
  dimension}},  Phys. Rev. Lett. {\bf 83} (1999) 3370--3373,
  [\href{http://arxiv.org/abs/hep-ph/9905221}{{\tt hep-ph/9905221}}].

\bibitem{SL-book}
V.~A. Marchenko, {\em Sturm-Liouville Operators and Applications}.
\newblock Birkhauser Verlag, CHE, 1986.

\bibitem{Chivukula:2022kju}
R.~S. Chivukula, E.~H. Simmons, and X.~Wang, {\it {Supersymmetry and sum rules
  in the Goldberger-Wise model}},  Phys. Rev. D {\bf 106} (2022), no.~3 035026,
  [\href{http://arxiv.org/abs/2207.02887}{{\tt arXiv:2207.02887}}].

\end{thebibliography}\endgroup

\end{document}